\documentclass[acmtog]{acmart}
\acmSubmissionID{861}

\usepackage{booktabs} %

\usepackage[ruled]{algorithm2e} %

\SetAlFnt{\small}
\SetAlCapFnt{\small}
\SetAlCapNameFnt{\small}
\SetAlCapHSkip{0pt}
\usepackage{cleveref}
\usepackage{xspace}
\usepackage{wrapfig}
\usepackage{soul}

\newcommand{\com}[1]{}

\definecolor{darkred}{rgb}{0.6, 0.1, 0.05}
\definecolor{blueish}{rgb}{0.0, 0.3, .6}

\acmJournal{TOG}

\setcopyright{acmcopyright}

\acmConference[SIGGRAPH Conference Papers'26]{Special Interest Group on Computer Graphics and Interactive Techniques Conference Papers}{July 19--23, 2026}{Los Angeles, CA, USA}

\acmBooktitle{Special Interest Group on Computer Graphics and Interactive Techniques Conference Papers, July 19-23, 2026, Los Angeles, CA, USA}

\begin{document}
\title{ADS: Random Sampling of Occupancy Functions using Adaptive Delaunay Scaffolding}

\author{Suzuran Takikawa}
\affiliation{
	\institution{University of British Columbia}
	\country{Canada}
}

\author{Leo Foord-Kelcey}
\affiliation{
	\institution{University of British Columbia}
	\country{Canada}
}

\author{Oliver Oxford}
\affiliation{
	\institution{University of British Columbia}
	\country{Canada}
}

\author{Nicholas Vining}
\affiliation{%
	\institution{NVIDIA}
	\country{Canada}
}
\affiliation{
	\institution{University of British Columbia}
	\country{Canada}
}

\author{Alla Sheffer}
\affiliation{
	\institution{University of British Columbia}
	\country{Canada}
}

\renewcommand\shortauthors{Takikawa et al.}

\begin{abstract}
Dense random sampling and surfacing of shapes encoded via implicit occupancy functions (OFs) are critical elements of many applications. Existing methods largely provide either one or the other of random sampling or mesh surfaces: ray shooting approaches deliver random samples with no connectivity, and grid-based methods deliver mesh surfaces but their sampling is highly biased.
We propose a new method which delivers both pseudo-random OF surface samples and an isosurface mesh connecting them. Our method achieves these goals while requiring an order of magnitude fewer function evaluations than prior approaches.
Key to our {\bf A}daptive {\bf D}elaunay {\bf S}ampling (ADS) approach is a progressively computed Delaunay tetrahedralization of points in 3D space, which we use as a sampling and surfacing scaffold. Starting from an initial coarse Delaunay scaffold, we repeatedly refine {\em crossing edges}, ones whose end vertices lie on opposite sides of the surface, augmenting the scaffold with points closer and closer to the surface. Each refinement step uses the Delaunay criterion to incorporate the newly added vertices into the scaffold, introducing new crossing edges.
We use the intersections of fine crossing edges with the OF surface as the output samples, and use the marching tetrahedra method to surface these samples. We subsequently use normal estimation to densify the sampling near fine features and in areas of high surface curvature. 
We validate ADS by sampling 150 inputs at different resolutions, and provide extensive comparisons to existing alternatives. Our experiments demonstrate significant improvement in accuracy/function evaluation count trade-off, and showcase downstream applications.
\end{abstract}

\newenvironment{parWithWrapFigure} %
{\begingroup
\setlength{\columnsep}{1em}%
\setlength{\intextsep}{0em}%
\setlength{\arraycolsep}{0pt}} %
{
\endgroup
}


\begin{teaserfigure}
 \centering
 \includegraphics[width=\linewidth]{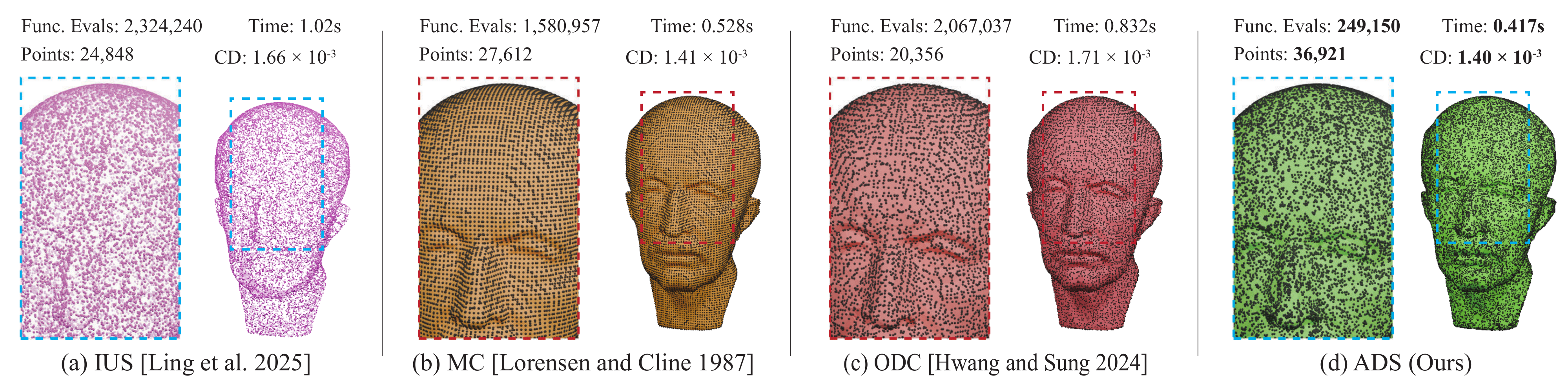}
  \caption{Ray-shooting based occupancy surface sampling \cite{ling2025uniform} (a) produces random samples, but does not recover sample connectivity. Grid-based methods \cite{lorensen1998marching,hwang2024odc} (c,d) recover iso-surface connectivity, but produce grid-biased sampling patterns (see zooms). Both families require large numbers of function evaluations to accurately approximate the inputs. ADS produces connected iso-surfaces with pseudo-random point placement and requires an order of magnitude fewer function evaluations to achieve similar accuracy, leading to faster runtimes. (Accuracy is reported by chamfer distance, or CD, above.)}
  \label{fig:teaser}
\end{teaserfigure}

\maketitle

\begin{figure*}
\centering \includegraphics[width=\linewidth]{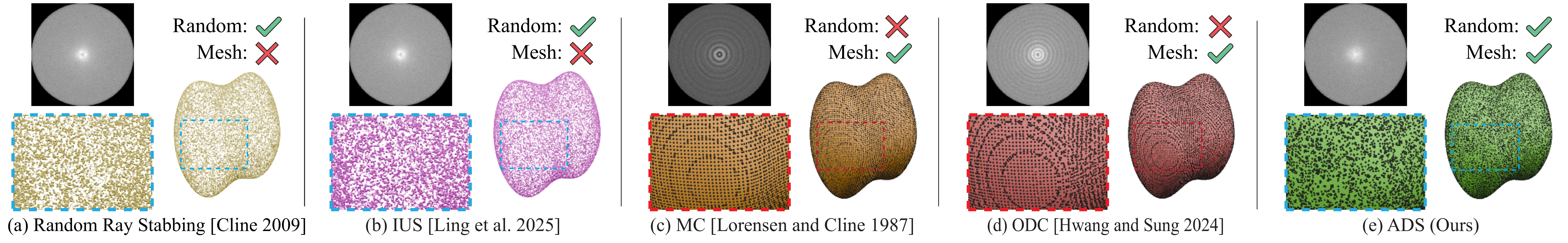}
\caption{Random \cite{cline2009dart} (a) and uniformly random \cite{ling2025uniform} (b) ray-casting provide provably random sampling of occupancy surfaces but generate no connectivity information. Grid based isosurfacing methods such as Marching Cubes \cite{lorensen1998marching} (c) or Occupancy dual contouring \cite{hwang2024odc} (d) output fully connected but grid-biased samplings. ADS generates pseudo-random samples and an isosurface connecting them (e). Inset spectral images highlight these differences: grid-based methods exhibit structured artifacts, while ADS and uniform sampling are largely artifact-free.}
\label{fig:other}
\vspace{-3mm}
\end{figure*}

\section{Introduction}
\label{sec:intro}
Recent years have seen a dramatic increase in the use of implicit functions as a 3D geometry representation. This increase is largely driven by the ease with which implicits can be encoded in neural form \cite{mescheder2019occupancy,chen2019learning,zhang2024nesi,takikawa2022dataset}. Occupancy functions (OFs) provide a particularly compact way to encode geometry, making them especially appealing for applications where memory is a major bottleneck. Processing occupancy and other implicit functions often requires densely and randomly sampling the surface they describe or, better yet, extracting a surface mesh connecting such samples. Ray-shooting methods successfully generate dense random samplings of implicit surfaces \cite{ling2025uniform} but do not recover the surface connectivity between them (Fig. ~\ref{fig:teaser}a, Fig.~\ref{fig:other}ab). Grid-based approaches \cite{lorensen1998marching,hwang2024odc} can be used to surface OFs, but the surface points or {\em samples} they generate are not randomly distributed and exhibit a clear grid bias (Fig.~\ref{fig:teaser}bc, Fig.~\ref{fig:other}cd). Using classical Delaunay refinement methods \cite{boissonnatoudot2005,oudot_roundtable_2005} for meshing OF surfaces become prohibitively time consuming in our setting (Sec.~\ref{sec:results}) requiring 36 minutes to sample 22K points on the example in Fig.~\ref{fig:teaser}, making them unsuitable for the applications we target.
We propose {\em Adaptive Delaunay Sampling} (ADS), a new fast and practical method for generating pseudo-random surface samplings and isosurface meshes connecting these samples (Fig.~\ref{fig:teaser}d, Fig.~\ref{fig:other}e).

A core challenge in sampling OFs is the lack of a way to compute the distance to the isosurface, or even a reliable function providing gradient information that can be used to guide the sampling (Sec.~\ref{sec:related}). Consequently, existing OF sampling and surfacing methods typically require huge numbers of function queries to obtain accurate OF surface samplings and/or meshes.  ADS is capable of achieving a similar degree of accuracy while using drastically fewer function evaluations than prior methods (Fig.~\ref{fig:teaser}, Sec.~\ref{sec:results}).  

\begin{figure}
\centering \includegraphics[width=\linewidth]{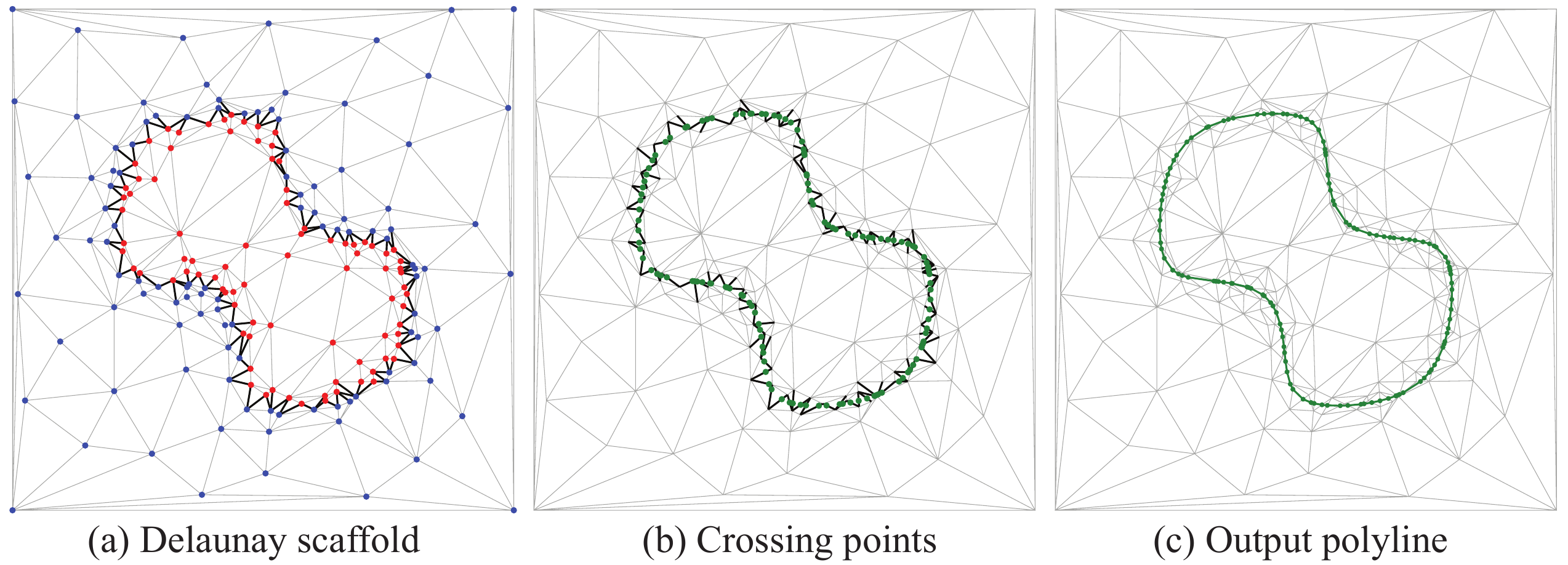}
\caption{2D illustration of our method: Delaunay scaffold with vertices concentrated around the target surface (a), inside vertices in red and outside in blue, crossing edges bolded.  (b) Crossing points (green) accurately capture the input shape  (c). Mesh (2D polyline) connecting the samples computed using marching tetrahedra (triangles).}
\label{fig:illustration}
\vspace{-6mm}
\end{figure}

At the core of our ADS method is the observation that a key step in sampling occupancy surfaces is computing {\em crossing edges}, namely line segments that connect pairs of vertices in 3D space where one vertex is inside the surface and one outside. Each such edge contains at least one {\em crossing point} which lies on the target surface. 
Given a crossing edge, one can progressively refine it to locate these crossing points (up to a given accuracy). The challenge of sampling occupancy surfaces can be therefore seen as one of creating a large set of well-distributed crossing edges whose crossing points densely cover the input shape. We achieve this goal by constructing a tetrahedral Delaunay scaffold spanning the function domain whose vertices are concentrated in the vicinity of the approximated surface, and densely sample the space surrounding it both inside and outside (Fig. ~\ref{fig:illustration}a). 
By construction, this scaffold contains a large dense and well-distributed set of crossing edges connecting inside and outside vertices (Sec. ~\ref{sec:overview}). The crossing points along these edges provide our desired sampling (Fig.~\ref{fig:illustration}b). Furthermore, the scaffold structure allows immediate computation of a mesh connecting the crossing points along these edges, using a variant of the marching tetrahedra method \cite{doi1991efficient} (Fig.~\ref{fig:illustration}c).  Our key challenge therefore becomes efficiently generating this scaffold, and specifically generating scaffold vertices in the region surrounding the target surface. Since we operate on OFs, we have no way to assess how close a vertex is to the surface; this limitation makes such placement particularly challenging.  

\paragraph{Overview.}
We generate the desired scaffolds using an incremental construction process, which starts by generating a well-spaced set of vertices within the bounding domain of the input surface, and then finding the simplicial Delaunay mesh of these points. We then progressively refine crossing edges, adding all newly introduced vertices to the Delaunay simplicial complex. The refinement substantially increases the number of crossing edges as these new vertices are connected to nearby inside and outside ones. Furthermore, maintaining the Delaunay property implicitly reduces the length of all crossing edges. In our experiments, each refinement iteration increases the number of crossing-edges by factor three on average, while decreasing the average crossing edge length by $\sim40\%$ in early iterations, with slowing decreases in later iterations. Once all crossing edges are sufficiently short, ADS computes the crossing points along them and meshes these points using a variation of the marching tetraheda method \cite{doi1991efficient}. ADS supports curvature adaptive sampling by introducing additional scaffold vertices in regions where adjacent sample points have significantly varying normals. This step facilitates effective sampling of fine geometry details.

We validate ADS by applying it to 150 diversely-sourced occupancy functions at different sampling rates, and compare the results against those of key alternatives. ADS produces dense random samples that capture the input geometry to a desired accuracy and does so with approximately 6.5 function evaluations per surface sample.  In contrast to all prior methods ADS supports adaptive sampling enabling denser sample density in high curvature regions.  Most importantly, ADS computes both random samples and their on-surface connectivity, something none of the prior approaches are capable of doing.  ADS is simple to implement and we will release our code for ease of use.

Our approach can be effectively extended to surfaces represented by implicit functions that encode information beyond occupancy, such as upper or lower bounds on distance to the encoded surface. In all these cases, our method can be further sped up by drastically reducing the number of function evaluations needed (Sec.~\ref{sec:results}). 

\section{Background and Related Work}
\label{sec:related}

\paragraph{Occupancy Function Surfaces}
Our work is motivated by the recent interest in generative methods for encoding and synthesizing content as neural occupancy functions, such as IM-Net \cite{chen2019learning}, occupancy networks \cite{mescheder2019occupancy}, and NESIs \cite{zhang2024nesi}. Other examples of occupancy functions include winding number functions computed on polygon soups \cite{barill2018fast} as well as shapes created via traditional CSG operations on varying primitives \cite{wyvill1998blob}. Manipulating the outputs of these methods often requires sampling or surfacing.

Implicit representations describe a closed surface $S$ as a zero-level-set of a function $F: \mathbb{R}^3 \rightarrow \mathbb{R}$; that is, the set of points in $\mathbb{R}^3$ whose value is a constant, usually zero. An occupancy function $\phi: \mathbb{R}^3\rightarrow{-1,1}$ represents a shape $S$ as the {\em boundary} of two discrete labels, $-1$  for outside and $1$ for inside. The distance to the shape $S$ at a given point is {\em a priori} unknown, and gradient information is likewise unavailable (finite differencing of occupancy functions to extract gradients is noisy and unreliable \cite{tornberg2002multi}). Moreover, with this definition even assessing if a point is {\em on} the surface is non-trivial - commonly points are deemed as on the surface if they lie on a sufficiently short line-segment whose end-points have opposite labels. Theoretical occupancy functions can be defined using three labels $-1$ for outside, $1$ for inside, and $0$ for on the surface, however due to numerical precision limitations sampling exact $0$ values only is impractical. Due to these challenges sampling and surfacing occupancy functions remains a hard, open problem. 

\paragraph{Sampling Implicit Surfaces}
There exist a range of approaches for sampling implicit functions. Strategies like rejection sampling \cite{dippe1985antialiasing,cline2009dart} require extensive oversampling of the implicit function to generate results. Particle system based approaches \cite{witkin1994using,levet2007marchingparticles} formulate implicit surface sampling as a particle repulsion problem and require repeatedly solving extremely large $n \times n$ linear systems, where $n$ is the number of samples being sought. It is not clear how to apply these methods to occupancy functions, since as mentioned above, one cannot reliably assess if a stand-alone point is on the surface. 

Many methods address the problem of sampling {\em signed distance functions} (SDFs), in which $F$ explicitly returns the distance to $S$. As exact SDFs satisfy a Lipschitz condition, they can be efficiently evaluated by methods such as sphere tracing \cite{hart1996sphere}. Occupancy functions have no reliable gradients and may not satisfy the Lipschitz condition. Wang et al. ~\shortcite{yifan2021iso} present a method for uniform sampling of implicit surfaces using a Newton iteration scheme. Their approach requires the existence of function gradients from which Jacobians are derived; as OFs often do not have reliable gradients, derivation of the Jacobian is not possible. 

Achieving a sampling of a surface with a specific noise spectral frequency (i.e. white or blue noise) is important for many tasks in computer graphics, as the efficacy of downstream applications often depends on the distribution of underlying samples \cite{wei2011differential,Singh2019Analysis}. Rejection sampling methods \cite{cline2009dart,chiu2022thesis,yuksel2015sample} start from a random set of sample points and downsample it to satisfy a desired frequency. ADS generates pseudo-random samplings, which can be easily downsampled to perfect white or blue noise, see supplementary for detailed discussion.

\paragraph{Sampling Occupancy Function Surfaces}
A common practical approach for sampling occupancy functions is to generate a large set of randomly oriented infinite lines within the parameter domain of the functions and use their intersections with the surface as samples \cite{cline2009dart,ling2025uniform}.  Ling et al. ~\shortcite{ling2025uniform} propose a strategy for computing such lines that generates a uniformly random point sampling of the surfaces. Our method achieves comparable or better approximation accuracy using twelve times fewer function evaluations than ray-casting and is two to three times faster (Fig.~\ref{fig:teaser}, Fig.~\ref{fig:other}, Sec.~\ref{sec:results}) and produces both samples and an isosurface connecting them.

Sharp and Jacobson \shortcite{sharp2022spelunking} accelerate sampling queries using interval arithmetic; they assume that the occupancy function is encoded as a neural network and require significant internal information about the network's construction in order to operate. Liu et al. ~\shortcite{liu2024neural} extend this idea to neural bounding volume computation. Stippel et al. \shortcite{stippel2025marching}'s 'marching neurons' method extends the same idea to a marching cubes type scheme. Our method operates on arbitrary occupancy functions and requires no {\em a priori} knowledge about the underlying encoding, as such it can be applied as-is to differently sourced inputs (Sec.~\ref{sec:results}). 

\paragraph{Isosurface Reconstruction for Implicit/Occupancy Surfaces}
Isosurfacing of implicit and occupancy functions is most commonly done using spatial subdivision approaches such as the classical Marching Cubes (MC)  algorithm \cite{lorensen1998marching}. These methods divide the volume the function is defined over into cells, locate cells containing crossing edges and for each cell containing one or more crossing edges generate per-cell triangulated surface patches connecting the crossing points along these edges and designed so as to jointly form a watertight isosurface. While other isosurfacing approaches had been explored in the past (see \cite{de2015survey} for a comprehensive survey) they fail to capture complex shapes and are thus rarely used in practice. 

Spatial subdivision approaches most commonly operate on hexahedral grids \cite{lorensen1998marching,gibson1998constrained,JuDual,hwang2024odc}.  The main difference between these methods is the quality of the triangulations they produce, as well as the number of surface samples utilized. Most recently,
\cite{hwang2024odc} proposed Occupancy Dual Contouring (ODC) a dual contouring approach specifically targeting OFs. They note that grid-based methods such as Marching Cubes often produce staircase-like outputs, and compensate for this using a hierarchical search pattern along cube faces which is later meshed using a dual contouring approach. The sampling patterns all these methods introduce exhibit clear axial bias, most pronounced in traditional MC outputs (Fig.~\ref{fig:other}c) but still distinctly notable in outputs of more recent methods such as ODC (Fig.~\ref{fig:other}d).  These methods require evaluating the surfaced functions at all grid points, and then performing additional evaluations along crossing edges to locate crossing points, and can become very costly when high approximation accuracy is required.  ADS requires an order of magnitude fewer evaluations than the above methods to generate random, same or higher accuracy outputs (Fig.~\ref{fig:teaser}, Sec. ~\ref{sec:results}).  

Early research proposed to reduce the number of evaluations and speed-up isosurfacing by using hierarchical octrees \cite{bloomenthal1988polygonization} instead of grids. Unfortunately, while grid-based isosurfacing can be easily parallelized, octree based approaches require recursive, and thus hard to parallelize, computation and are thus ill-suited for the modern GPU-heavy processing pipelines. 

Spatial subdivision can be applied to tetrahedral meshes, using the marching tetrahedra (MT) isosurfacing scheme \cite{doi1991efficient}. Existing approaches utilize regular tetrahedral latices, introducing similar bias to that of grid-based methods. Muller et. al \shortcite{Mller1997VisualizationOI} propose a hierarchical method that mimics the octree isosurfacing approach, and subdivides crossing tetrahedra. This approach employs a similar recursive mechanisms to the the octree methods making it similarly ill-suited for modern processing pipelines.  An appealing theoretical approach to produce unbiased samples and interpolating isosurfaces could be to generate a Delaunay tetrahedralizaiton of a dense set of randomly sampled points sampled in the occupancy function domain. Unfortunately, achieving accurate approximation requires on the order of one to ten million samples, and introduces a much larger number of crossing edges than standard MC. In our experiments, sampling and isosurfacing such dense Delaunay meshes took fifteen to twenty times longer than ADS to generate comparable accuracy samplings. 

Several approaches are specifically designed for reconstructing occupancy and other implicit surfaces that are defined via piece-wise linear interpolation of values specified on a regular grid or octree \cite{ju2006intersection}.  These methods leverage the expectation that the surface intersects the input grid or octree edges at most once. We target the general case and make no assumptions about the sourcing of the occupancy functions. 

\paragraph{Delaunay Refinement Methods}
Classical Delaunay refinement methods for surface meshing \cite{oudot_roundtable_2005,boissonnatoudot2005} insert surface sample points into a Delaunay tetrahedralization sequentially, checking at each step whether new edges cross the surface. Densely sampling occupancy functions using these methods requires extremely large numbers of intersection tests. Since each insertion depends on prior ones, intersection tests cannot be batched across steps, precluding GPU-parallel occupancy evaluation. These methods are thus orders of magnitude slower than our method and the baselines we compare against \cite{ling2025uniform, lorensen1998marching, hwang2024odc, cline2009dart} (Sec.~\ref{sec:results}).

\paragraph{Delaunay Surfacing Methods}
Finally, we note that a class of methods exists that use Delaunay triangulations/tetrahedralizations to {\em reconstruct} surfaces given a set of sample points as input~\cite{amenta1998surface,amenta2001power,alliez2005variational,kolluri2004spectral} These should not be confused with our method, which {\em generates} the sample points given an occupancy function and uses a Delaunay scaffold to facilitate their placement. Xu et al. ~\shortcite{xu2011capacity} gives a method for sampling the surface of an already-known polygon in 2D using an iteratively updated capacity-constrained Delaunay triangulation. Their method does not propose a 3D extension.

\begin{figure*}
\centering \includegraphics[width=\linewidth]{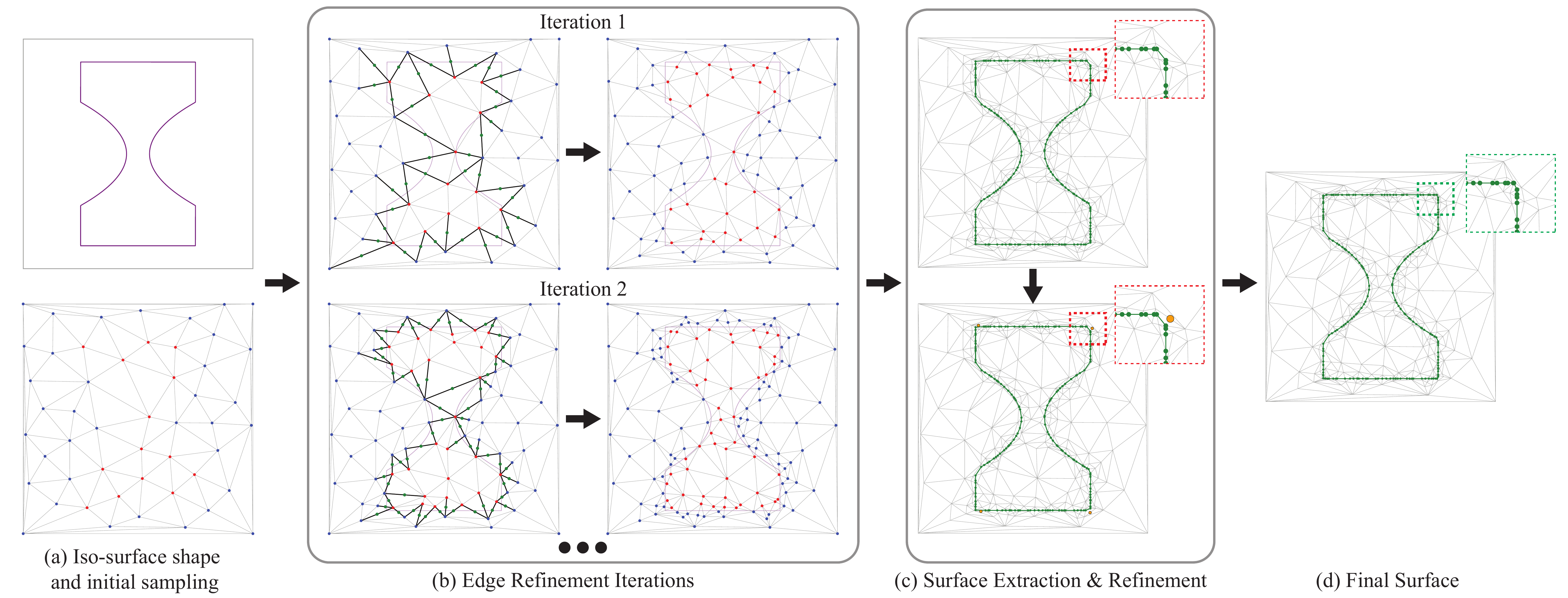}
\caption{ADS overview (in 2D). Left to right: (a) we sample a coarse, well-spaced initial point set within the bounding domain using Poisson disc sampling, then construct its Delaunay triangulation. (b) We iteratively refine the scaffold by identifying crossing edges, and inserting new vertices along them. (c) The surface is extracted using marching tetrahedra, and locally refined to capture thin or high-curvature iso-surface details; (d) the final output iso-surface and samples.}
\label{fig:overview}
\end{figure*}

\section{Method}
\label{sec:alg}

\subsection{Overview}
\label{sec:overview}

Given an occupancy function $\phi: \mathbb{R}^3 \rightarrow \{-1,1\}$ defined over a bounded domain $\Omega \subset \mathbb{R}^3$, our goal is to generate a dense set of points on the surface $S$ encoded by $\phi$ that accurately approximates it, as well as a triangle mesh interpolating these samples and approximating $S$.  We aim to do so as efficiently as possible, and in particular to minimize the number of function evaluations necessary. Since occupancy functions do not allow directly evaluating if a point is on the surface, methods for sampling such surfaces rely on {\em crossing edges}, line segments connecting a vertex inside the surface $S$ to a vertex outside it. Each crossing edge intersects $S$ at least once, and a crossing point can be located to precision $\epsilon$ via binary search using only $\mathcal{O}(\log(1/\epsilon))$ occupancy queries. The challenge of computing quality surface samples can therefore be recast as one of  generating a large set of well-distributed short crossing edges (the shorter the edge the faster the binary search). 

We generate the desired crossing edges by constructing and progressively refining a {\em Delaunay scaffold}, a tetrahedral mesh whose vertices concentrate near the target surface. This choice is guided by the following key observations. First, in a 3D Delaunay tetrahedralization, each vertex has an average of approximately 15 incident edges  \cite{meijering1953interface,Boissonnat2009Incremental}. Consequently, each vertex near the surface connects to many neighbors, dramatically increasing the likelihood that some of these neighbors lie on the opposite side of the surface; this contrasts with a regular hexahedral grid whose vertices are only connected to 6 neighbours. 
\begin{wrapfigure}[4]{l}{.525\columnwidth}
	\setlength{\columnsep}{0em}
	\setlength{\intextsep}{0em}
	\vspace*{-.175in}
	\includegraphics[width=.6\columnwidth]{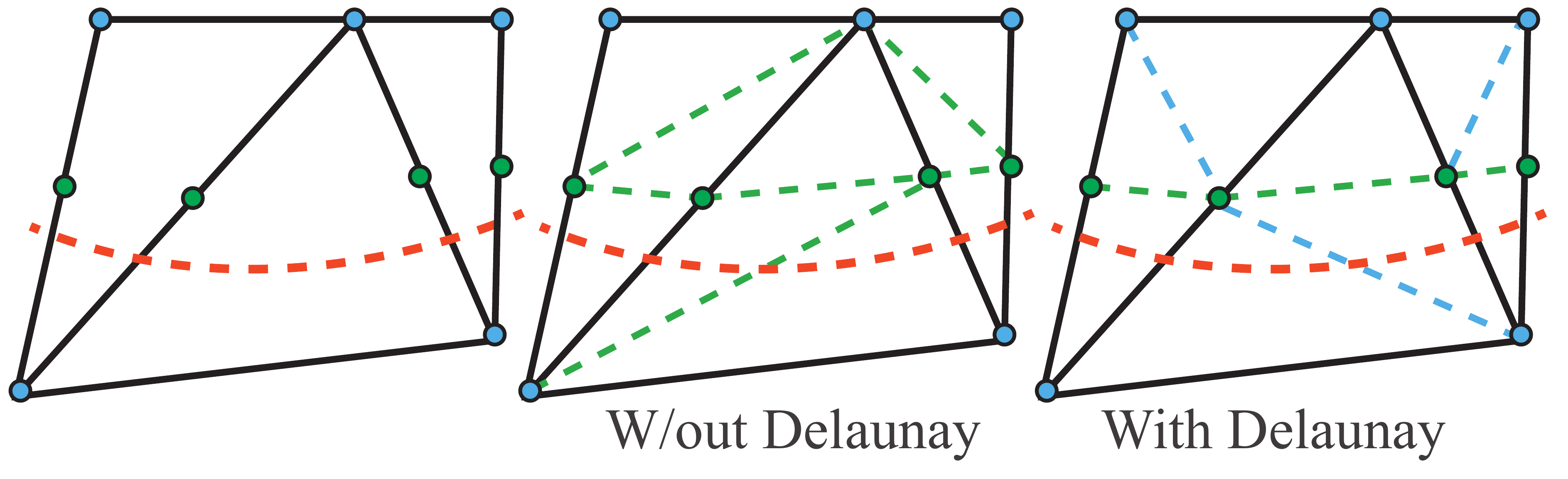}
\end{wrapfigure}
Second, by maintaining the Delaunay property post-insertion, the newly introduced crossing edges are on average much shorter than those introduced by the previous iteration (see inset). Note that while the Delaunay property is core to ADS, it performs adequately using scaffolds that are just approximately Delaunay. As such we can overcome co-planar configurations and slivers. Third, we observe that any scaffold tetrahedron that contains a crossing edge must, by definition, intersect the surface; and thus has either two or three other crossing edges. The intersection of the surface and the tetrahedron defines a local iso-surface patch, and the union of these patches defines a closed iso-surface mesh approximating the occupancy surface. This iso-surface interpolates our samples and can be efficiently extracted using Marching Tetrahedra \cite{doi1991efficient}.

\subsection{Method Pipeline}
\label{sec:pipeline}

Our scaffold computation strategy leverages the observations above and proceeds in four phases; see (Fig.~\ref{fig:overview}) for a 2D illustration.

\paragraph{Initialization.} 
We generate a coarse, well-spaced initial point set within the bounding domain using Poisson disc sampling \cite{bridson2007fast}, then construct its Delaunay tetrahedralization. We query $\phi$ at each vertex to determine its inside/outside classification. We keep this point set very coarse to avoid redundant function evaluations.

\paragraph{Edge Refinement.} 
We iteratively refine the scaffold by identifying crossing edges longer than a threshold $\tau$ and inserting new vertices along them (Sec.~\ref{sec:edge_refinement}). These new vertices are added to the Delaunay complex, which automatically connects them to nearby vertices. We query the OF $\phi$ at each new vertex to determine its inside/outside classification and detect crossing edges on the updated scaffold. Each refinement iteration increases the number of crossing edges by a factor of three to four on average. Once all crossing edges are shorter than $\tau$, we locate crossing points along them via binary search. The binary search is vectorized across all edges, using batch queries to $\phi$. This amortizes per-query overhead (e.g. GPU dispatch costs for neural network inference) and enables efficient GPU utilization when $\phi$ is implemented as a neural network. To avoid redundant computation, we track which edges are new and which existed in prior iterations. In each iteration, we only examine newly introduced edges, as all other edges retain their previous classification.
    
\paragraph{Surface Extraction.}
Following edge refinement, we extract a triangle mesh from the scaffold using a variant of the marching tetrahedra algorithm \cite{doi1991efficient}.  We note that our edge refinement process guarantees that all edges in the iso-surface mesh are shorter than $2\tau$ since each such edge connects points that lie on a scaffold triangle with two edges shorter than $\tau$. Triangle inequality hence dictates that these iso-surface edges are at most $2\tau$ long.  In practice since the scaffold triangles are well shaped most iso-surface edges are shorter than $\tau$. 

\paragraph{Mesh-Guided Local Refinement}
We observe that while the previous steps produce surface samples which are at most $2\tau$ apart in Euclidean space, the sampling may nevertheless fail to capture thin or high-curvature iso-surface details (Fig.~\ref{fig:overview}c). Notably, such situations can be detected by analyzing vertex normals on the isosurface mesh. Specifically, large differences between normals at edge endpoints are often indicative of undersampling in the vicinity of those edges (Fig.~\ref{fig:overview}c, red).  Our mesh guided refinement step (Sec.~\ref{sec:mesh_refinement})  uses normal analysis to guide mesh refinement in undersampled areas. It introduces new scaffold vertices at strategically selected locations and repeats the edge refinement and surface extraction steps.

\subsection{Edge Refinement}
\label{sec:refinement}
\label{sec:edge_refinement}

\begin{figure}
\centering \includegraphics[width=\linewidth]{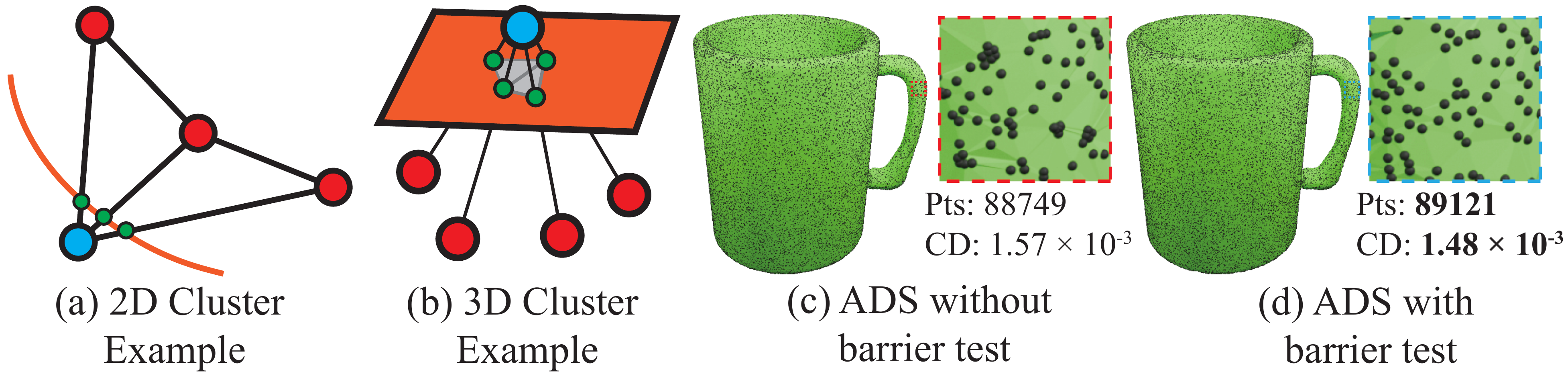}
\caption{Impact of barrier test: (a) 2D illustration: naive midpoint sampling may place vertices close to the surface resulting in surface samples being placed very close to one another. (b) The resulting artifacts are even more pronounced in 3D, where a single ``on surface'' vertex can introduce multiple highly clustered samples. (c,d) output samples without/with barrier test.}
\label{fig:barrier}
\vspace{-2mm}
\end{figure}

The refinement phase iteratively processes edges of the Delaunay scaffold, and identifies crossing edges that are longer than our spacing threshold $\tau$. We subdivide these edges by inserting new vertices. A naive subdivision approach would be to simply introduce the midpoints of the edges, as these are likely to be closer to the iso-surface than either of the endpoints. Using midpoint placement along all long crossing edges may, however, introduce vertices arbitrarily close to the iso-surface. Sample points placed on crossing edges attached to such vertices are going to be very close to these vertices, and in turn to one another (Fig.~\ref{fig:barrier}ab). 
\begin{wrapfigure}[6]{l}{.17\columnwidth}
	\setlength{\columnsep}{0em}
	\setlength{\intextsep}{0em}
	\vspace*{-.175in}
	\includegraphics[width=.25\columnwidth]{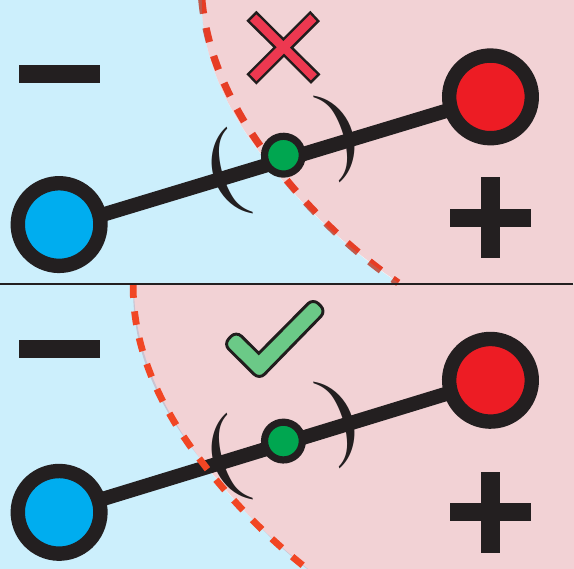}
\end{wrapfigure} Such samples are wasteful, as having more nearly co-located samples does not improve approximation accuracy, and degrade the randomness of the output sampling (Fig.~\ref{fig:barrier}bc). To reduce the prevalence of such ``on surface'' vertices, we use a barrier test. For each midpoint $m$ we consider adding, we query the OF $\phi$ to obtain the sign at points a small distance to the left and right of it, $m_l$ and $m_r$, along the edge (see inset). If the signs at the left and right points differ, this suggests that $m$ is very close to the surface. If the signs are the same we add the midpoint $m$ to the scaffold. Otherwise, instead of adding $m$ to the scaffold, we add a vertex along the edge  $1/3$  of the distance away from its end-vertex closer to the surface (a vertex is deemed closer if it is on the opposite side of the surface from the midpoint).  While this process does not fully prevent insertion of ``on surface'' points, as it does not capture cases where the crossing edge grazes the surface or where ``on surface'' points are introduced in other stages of the method, it dramatically reduces their prevalence.  We use a single batched function call to query the OF $\phi$ to determine the signs at all midpoints and their barrier endpoints. 

We insert the new  vertices incrementally into the Delaunay tetrahedralization \cite{bowyer1981computing,watson1981computing}, leveraging the original edge end vertices as insertion hints for efficient point location. We insert all new vertices from each iteration of refinement, regardless of whether their generating edges remain part of the Delaunay complex. Thus, insertion order has no effect. This process is extremely efficient, taking just 3 microseconds on average to insert each vertex.

\begin{figure}
\centering \includegraphics[width=\linewidth]{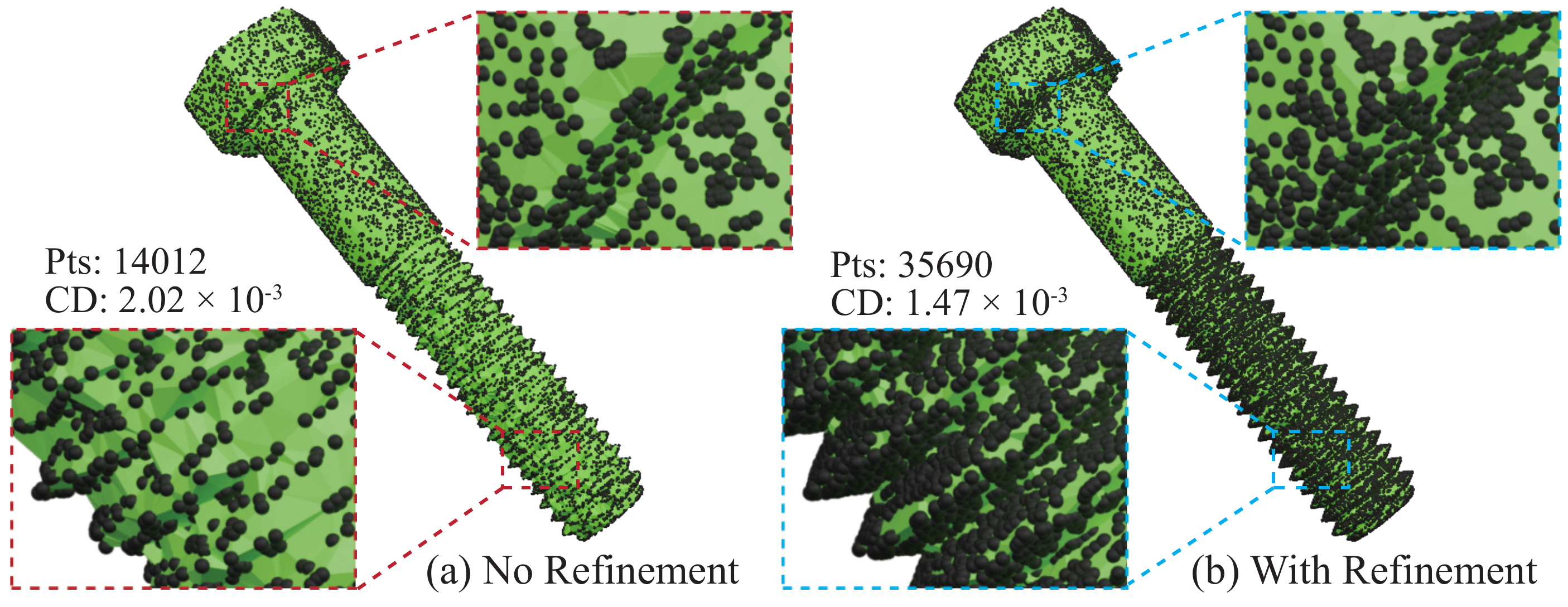}
\caption{Refinement: (a) initial sampling and iso-surface; (b) sampling and isosurface after refinement.}
\label{fig:refinement}
\vspace{-6mm}
\end{figure}

\subsection{Mesh-Guided Refinement}
\label{sec:curvature}
\label{sec:mesh_refinement}

Refining scaffold via crossing-edge splitting caps the surface distance between adjacent samples, but may not capture thin or high curvature features smaller than this cap (Fig.~\ref{fig:overview}c, Fig.~\ref{fig:refinement}a). Our refinement process addresses both scenarios simultaneously and automatically increases sampling rate across high-curvature surface regions. Our refinement is guided by surface normal variation. In particular, we note that on piece-wise smooth surfaces the normals at adjacent surface points are expected to be similar. The only exception is sharp features where normals on two sides of the feature sharply diverge. Accurately capturing such features benefits from dense sampling in their vicinity. We use these observations to refine the sampling when normals at samples connected by iso-surface mesh significantly diverge. We avoid oversampling next to sharp features by applying refinement only when the edges are sufficiently long. 

\paragraph{Normal Estimation.}
We estimate normal directions separately for each pair of sample points connected by an iso-surface mesh edge and compute each point's normal as the average of its incident isosurface triangle normals, {\em excluding} triangles shared by both points. Excluding the shared faces provides a more sensitive measure of local curvature: if the surface is locally flat, the remaining triangles would have similar normals, but if the edge lies on a crease or high-curvature region, the normal divergence will be more pronounced. 

\paragraph{Refinement Criteria.}
For each edge in the mesh, we compare the normals at its two end points and mark it for refinement if all of the following criteria are true: the angle between the normals exceeds our threshold, the edge is sufficiently long, and the edges in the scaffold tetrahedra containing the two isosurface triangles attached to the shared edge are sufficiently long. The latter criteria prevent over-refinement near sharp features. 

\paragraph{Refinement Point Placement.}
We then refine the tetrahedra containing the isosurface faces attached to the edge. These two tetrahedra intersect the iso-surface, and thus contain either three or two \begin{wrapfigure}[3]{l}{.22\columnwidth}
	\setlength{\columnsep}{0em}
	\setlength{\intextsep}{0em}
	\vspace*{-.22in}
	\includegraphics[width=.29\columnwidth]{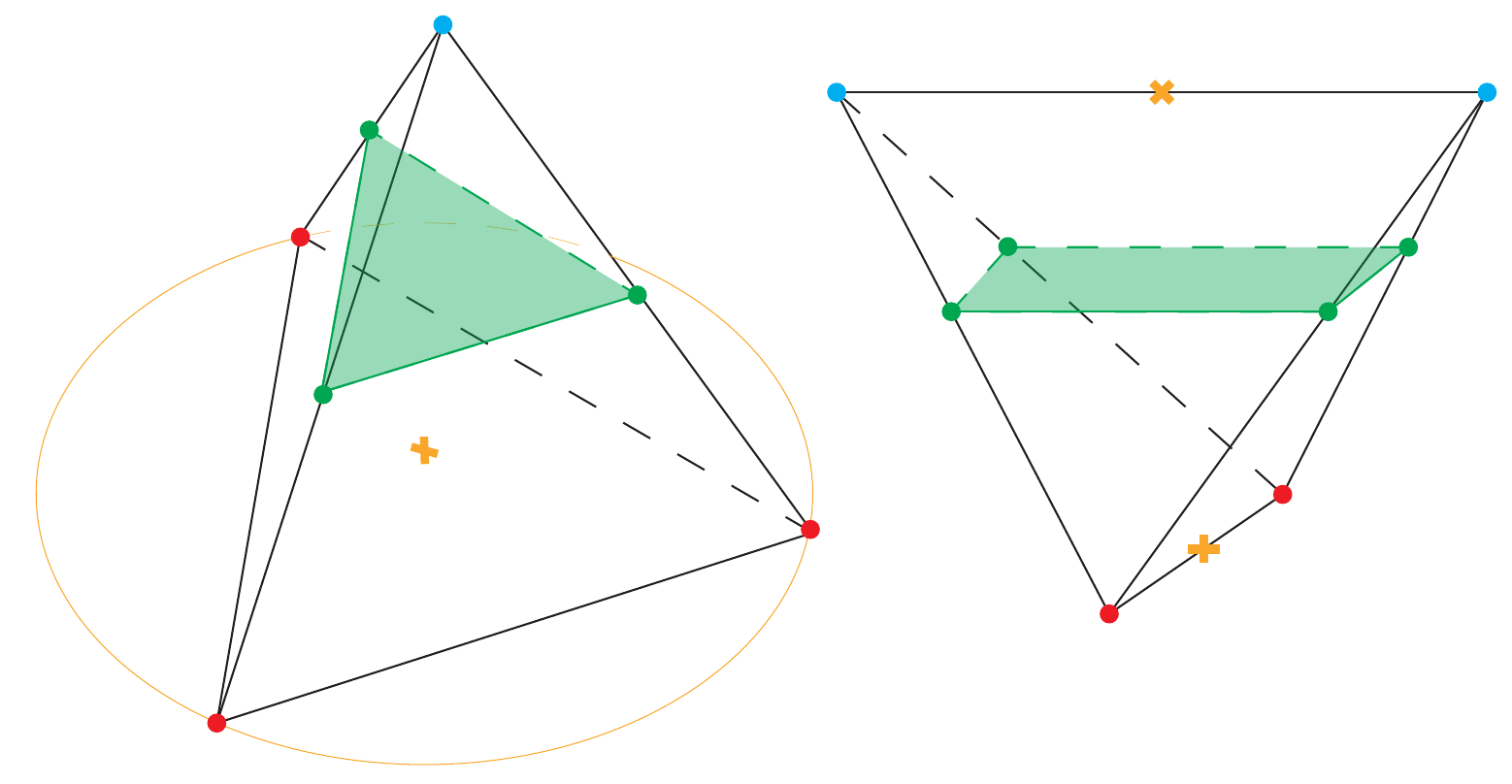}
\end{wrapfigure} 
crossing edges, (see inset, new vertex locations marked by X). In the first case, the tetrahedron has a triangular face that is entirely on one side of the input OF. In this case we insert a vertex at the circumcenter of this triangle. 
In the second case, the tetrahedron has one edge entirely inside the input OF and one outside; in this case we insert the midpoints of both of these edges.  

These new vertices are inserted into the Delaunay scaffold, and the algorithm repeats the edge refinement and meshing phases. We found that one round of refinement was sufficient to achieve accurate approximation on the inputs tested (Sec.~\ref{sec:results}). Additional iterations can be used to generate more strongly curvature adaptive sampling (Fig.~\ref{fig:curvature}).  Unless stated otherwise all results shown/reported were generated with one round of refinement.

\section{Results}
\label{sec:results}
\begin{figure*}
	\centering
	\includegraphics[width=\linewidth]{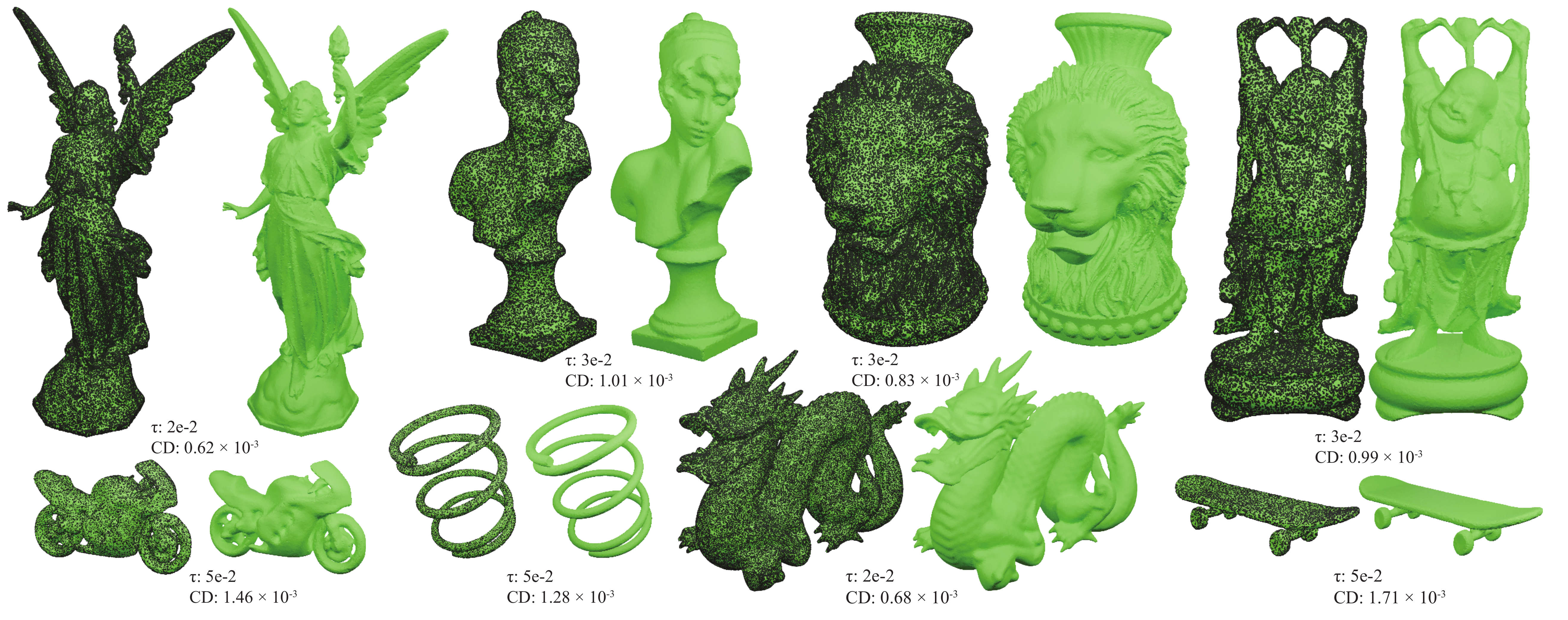}
	\caption{A gallery of our samplings and isosurface meshes using different sampling resolutions and input sources. Skateboard and motorcycle \cite{3DShape2VecSet}, spiral, Sapphos, happy Buddha \cite{zhang2024nesi}, Lucy, dragon, vaselion \cite{barill2018fast}. }
	\label{fig:ourgallery}
\end{figure*}

\begin{figure*}
	\centering \includegraphics[width=\linewidth]{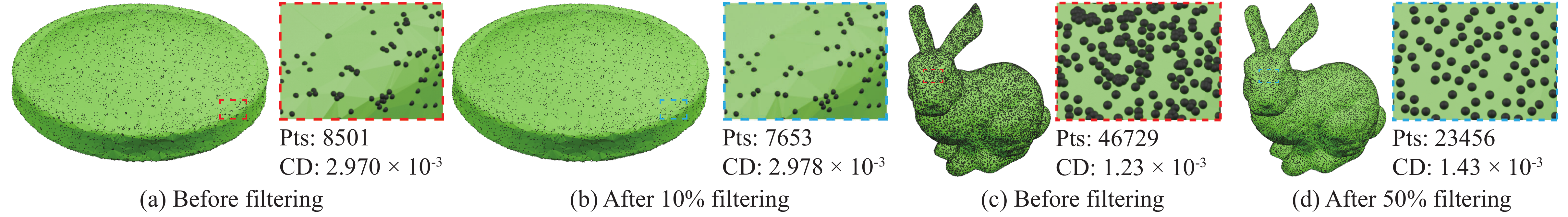}
	\caption{ADS outputs (a) can be directly filtered (rejection sampling plus edge collapse) using different rejection sampling thresholds: (b) 10\% filtering further reduces prevalence of close-by samples with minimal impact on accuracy (c) more aggressive (50\%) filtering produces blue noise sample distribution.}
	\label{fig:filtering}
	\label{fig:filter}
	\label{fig:noise}
\end{figure*}

We evaluate ADS on 150 inputs sourced from different shape sets and occupancy function sources: 50 inputs are neural occupancy functions learned from ShapeNet \cite{chang2015shapenet} data using Zhang et al.~\shortcite{3DShape2VecSet}; 50 are clamped winding number functions  \cite{barill2018fast} computed on inputs from \cite{myles2014robust}, and 50 are neural explicit function intersections (NESIs) \cite{zhang2024nesi} computed on inputs in the NESI dataset. These include both organic and CAD shapes and include many frequently used shapes (e.g. {\em max}, Fig.~\ref{fig:teaser}, {\em bunny}, Fig.~\ref{fig:filter}).  We sampled each input at three resolutions: $\tau=0.05,0.03,0.02$. See Fig.~\ref{fig:ourgallery} for a collection of representative outputs. We show 22 results throughout the paper, additional results included in the supplementary.  Unless explicitly ablated below, all other parameters were fixed across all experiments. See supplementary for evaluation details and parameter settings. 

\paragraph{Quantitative Evaluation}
We evaluate our outputs in terms of approximation quality and point distribution by measuring the distance between our samples and a dense, accurate sampling of the input surfaces (for the Myles \shortcite{myles2014robust} dataset where we expect the occupancy surface to strictly align with the mesh, we sample the meshes directly, for the other datasets we use ray shooting \cite{ling2025uniform} to generate 5 million points on the surface sampled with high accuracy). We compare our results against those of other methods at the same or worse $L^1$ chamfer distance, a standard comparison technique. On average the $L^1$ chamfer distances ($\times 1000$) between our outputs and the input surfaces were $1.46, 0.93, 0.67$ for $\tau=0.05,0.03,0.02$ respectively; all OFs defined over $[-1,1]^3$. Our evaluation counts range, on average, from 282K to 1.6M, and depend on sampling resolution and input surface complexity (Tab.~\ref{tab:comparison}). 

\paragraph{Spectral Properties} 
We analyze spectral properties using differential domain analysis~\cite{wei2011differential}, which expresses spectral behavior in terms of pairwise sample differentials and is closely related to auto-correlation. Our analysis suggests that the differential-domain spectrum of our samplings is very similar to that of white noise sampling, with the one difference arising from the undesirable sample clustering (Fig.~\ref{fig:filtering}a). Following even a small amount of Poisson disc filtering (Fig.~\ref{fig:filtering}b) the spectra become even more similar. This is not the case for MC or ODS, whose spectra exhibit clear bias even after filtering. See supplementary for details.

\paragraph{Performance} 
Our runtimes range from 0.65 seconds to 2.04 seconds on average and were measured on a machine with an AMD Ryzen 9 9900X CPU and a NVIDIA RTX 2080 GPU. On average we use 1.3GB of memory; the amount used ranges from 1 to 2 GB, depending on output mesh size. See supplementary for more details.

\paragraph{Ablations}
Figures~\ref{fig:barrier},~\ref{fig:refinement} ablate important algorithmic choices we make. Additional ablations are included in the supplementary.

\subsection{Comparisons}
 
\begin{figure*}
	\centering \includegraphics[width=\linewidth]{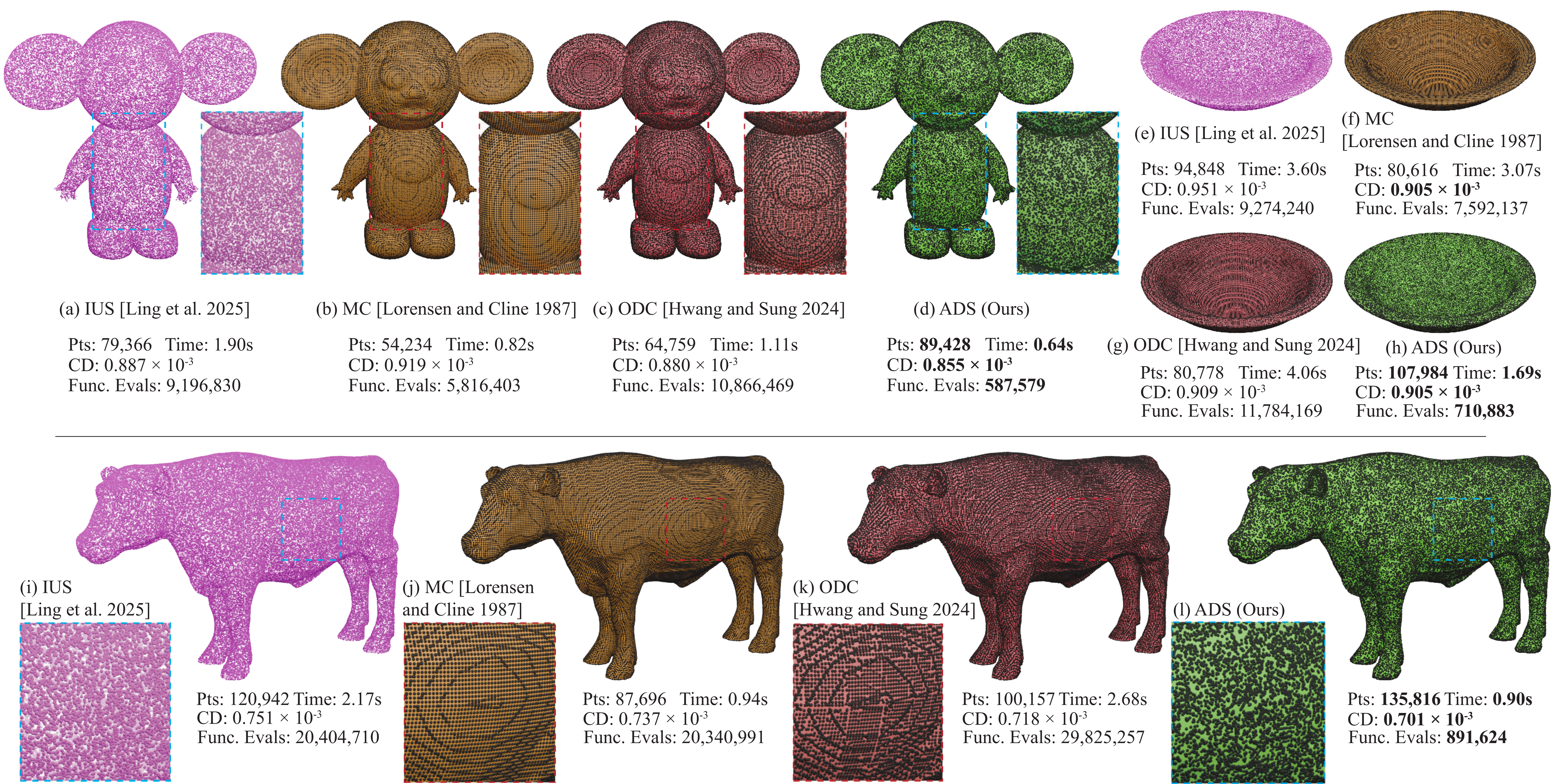}
	\caption{Additional visual comparisons of our ADS method to uniform ray casting \cite{ling2025uniform}, marching cubes (MC) \cite{lorensen1998marching}, and occupancy dual contouring (ODC) \cite{hwang2024odc}. We achieve higher accuracy and higher sample count with fewer occupancy function evaluations and less time. Contrary to \cite{ling2025uniform}, we generate both samples and iso-surfaces. Contrary to grid-based methods, we produce random, unbiased samplings.}
	\label{fig:gallery}
\end{figure*}

We compare our method to four representative alternatives: random ray casting, uniform ray casting \cite{ling2025uniform}, marching cubes (MC) \cite{lorensen1998marching}, and occupancy dual contouring (ODC) \cite{hwang2024odc}. We use the authors' implementation of uniform ray casting \cite{ling2025uniform}, occupancy dual contouring \cite{hwang2024odc} and our own implementation of the other two methods, since there is no standard optimized code available for either. As already noted, unlike ADS, the former two methods do not extract surfaces. As comparisons (Figs.~\ref{fig:teaser},~\ref{fig:other},~\ref{fig:gallery}) show, while our samplings are random, the ones produced by MC and ODC have prominent grid-based structure. We further quantitatively compare all methods in terms of the tradeoff between the number of occupancy function evaluations required, the accuracy of the sampling obtained, and the number of output surface samples (Tab.~\ref{tab:comparison}). Specifically, for each of our results we modify the main parameter of the other method (number of rays for ray-shooting approaches, grid-resolution for MC and ODC) so as to produce a result with maximally close, but still higher chamfer, and report the number of evaluations and runtime required to achieve this. We also report the number of surface samples produced.  
 
\begin{figure*}
\centering \includegraphics[width=0.89\linewidth]{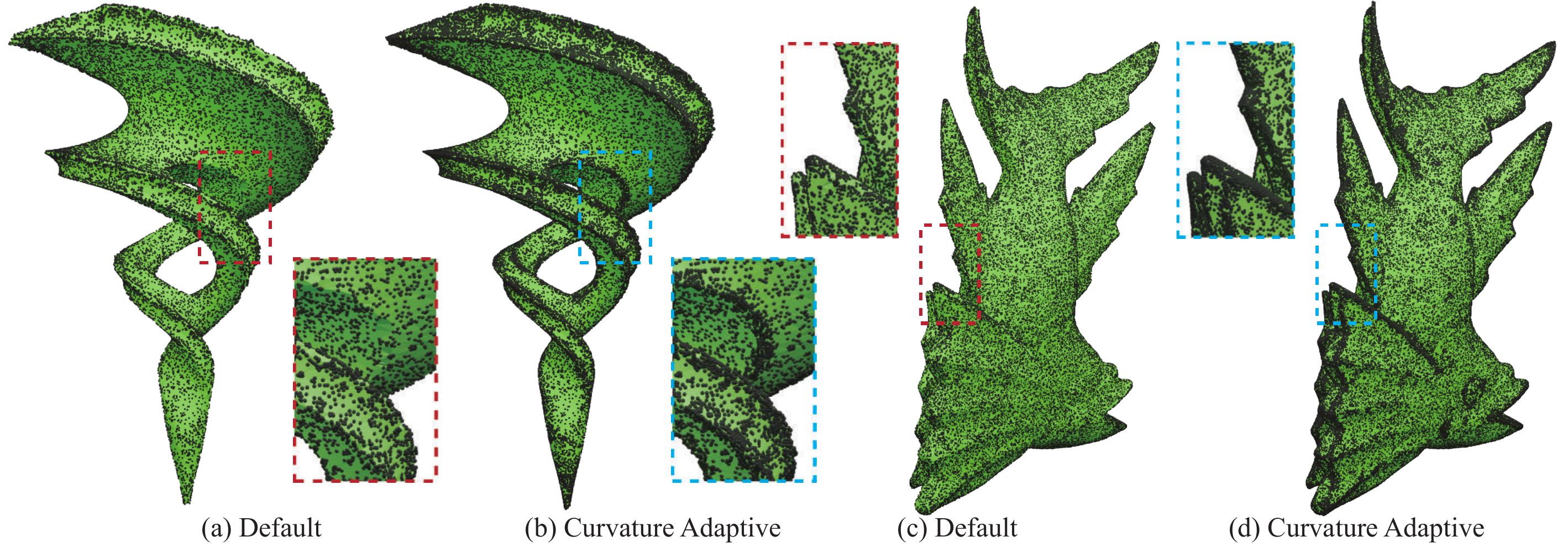}
	\caption{Curvature adaptive sampling: (ac) Standard ADS outputs; (bd) Strongly curvature adapted outputs with more samples in higher curvature areas.}   
	\label{fig:curvature}
\end{figure*}

As shown in the table, ADS achieves better accuracy (lower chamfer distance) while requiring 91\%-94\% fewer evaluations on average than the alternatives and is significantly faster.  Notably, ADS produces roughly 40\% more surface samples on average than the grid-based alternatives, making its outputs well suited for rejection sampling applications that seek to keep as many points as possible while achieving some desired distribution. This high sampling rate facilitates other downstream applications such as quality remeshing - enabling us to create meshes with well shaped triangles (Fig.~\ref{fig:meshing}). Contrary to all approaches ADS automatically adapts to surface curvature, producing more samples in higher curvature regions (Fig.~\ref{fig:curvature}ac), the adaptivity can be further increased (Fig.~\ref{fig:curvature}bd). None of the prior methods are capable of such adaptation. 

\paragraph{Comparison to Classical Delaunay Refinement.} 
We compare to the CGAL~\shortcite{cgal} implementation of Boissonnat and Oudot~\shortcite{boissonnatoudot2005}, by applying their and our methods to three randomly selected inputs, one from each of our datasets, using comparable sampling density. For winding number OFs~\cite{barill2018fast}, CGAL required 17 seconds (6.8M evaluations, 10,829 samples, CD~1.99); ADS achieved same accuracy in 0.22 seconds using only 127K evaluations. The gap widens substantially for GPU-based neural OFs: for NESI~\cite{zhang2024nesi}, CGAL runtime was 1,260 sec. (21 min.) versus 0.23 seconds for ADS; for the neural OFs of~\cite{3DShape2VecSet}, CGAL took 8,968 sec. (2.5 hours) versus 0.71 sec. for ADS.

\begin{table}
	\scriptsize
	\setlength{\tabcolsep}{6pt}
	\centering
	\begin{tabular}{lccccc}
		\textbf{Method} & \textbf{Resolution}  & \textbf{Chamfer $\downarrow$} & \textbf{\# Output $\uparrow$} & \textbf{\# Evals $\downarrow$} & \textbf{Time (s) $\downarrow$} \\
		 &  & \textbf{$\times 10^{3}$} & \textbf{\# Samples$\uparrow$} &  &  \\
		\hline
		RRS & low& 1.81 & 26,124 & 3,005,289 & 0.70 \\
		RRS & medium & 0.98 & 96,299 & 11,041,497 & 2.55 \\
		RRS & high & 0.69 & 219,909 & 25,387,544 & 5.81 \\
		RRS & all  & 1.16 & 114,111 & 13,144,777 & 3.02 \\
		\hline
		IUS & low& 1.77 & 26,471 & 2,510,734 & 0.65 \\
		IUS & medium & 0.97 & 100,132 & 9,564,659 & 2.44 \\
		IUS & high & 0.69 & 226,502 & 21,665,846 & 5.37 \\
		IUS & all  & 1.14 & 118,109 & 11,286,081 & 2.83 \\
		\hline
		ODC & low & 1.77 & 22,021 & 2,453,094 & 0.88 \\
		ODC & medium & 0.99 & 73,172 & 11,088,175 & 2.34 \\
		ODC & high & 0.69 & 162,438 & 32,240,003 & 5.92 \\
		ODC & all  & 1.15 & 85,877 & 15,260,424 & 3.04 \\
		\hline
		MC & low & 1.53 & 26,182 & 1,611,151 & \textbf{0.39} \\
		MC & medium & 0.97 & 69,369 & 6,766,395 & 1.45 \\
		MC & high & 0.69 & 155,607 & 22,106,636 & 4.77 \\
		MC & all  & 1.06 & 83,719 & 10,161,394 & 2.20 \\
		\hline
		ADS & low & \textbf{1.46} & \textbf{43,038} & \textbf{281,646} & 0.65 \\
		ADS & medium & \textbf{0.93} & \textbf{111,997} & \textbf{733,440} & \textbf{1.07} \\
		ADS & high & \textbf{0.67} & \textbf{244,889} & \textbf{1,604,561} & \textbf{2.04} \\
		ADS & all  & \textbf{1.02} & \textbf{133,308} & \textbf{873,215} & \textbf{1.25} \\
	\end{tabular}
	\caption{Quantitative comparison of ADS (bottom) against prior art (random ray stabbing (RRS) \cite{cline2009dart}, uniform ray casting (IUS) \cite{ling2025uniform}, marching cubes (MC) \cite{lorensen1998marching}, and occupancy dual contouring (ODC) \cite{hwang2024odc}) across different resolutions (low, medium, high, $\tau=0.05,0.03,0.02$ respectively) and datasets. ADS consistently achieves better accuracy (lower chamfer) and generates more samples while requiring less function evaluations. It is faster than RRS, IUS and ODC and becomes faster than MC at higher sampling resolution.} 
	\label{tab:comparison}
	\vspace{-6mm} 
\end{table}

\subsection{Extensions and Applications}

\paragraph{Rejection Subsampling.} As Fig.~\ref{fig:noise} shows our samplings serves as a good initialization for generating blue noise via rejection sampling \cite{yuksel2015sample}. Minimal downsampling (Fig.~\ref{fig:noise}b) improves point distribution with negligible reduction in accuracy. More aggressive downsampling produces blue noise samples which continue to well approximate the input. 

\begin{figure*}
	\centering \includegraphics[width=\linewidth]{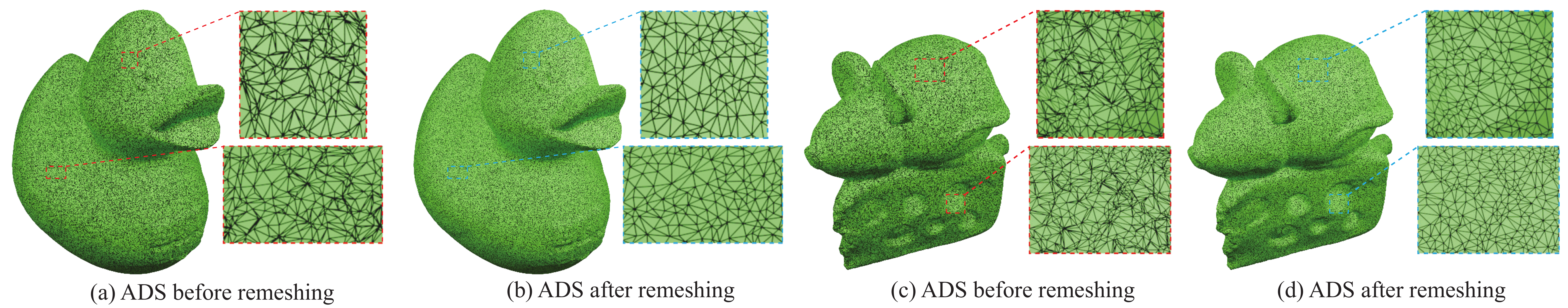}
	\caption{Optimizing our iso-surface meshes by collapsing short edges and flipping edges to improve smoothness and aspect ratio: (ac) initial iso-surface meshes. (bd) Optimized meshes.} 
	\label{fig:remeshing}
	\label{fig:meshing}
\end{figure*}

\paragraph{Mesh Optimization.} 
Our output isosurface meshes can be quickly improved through standard edge collapse and flip operations, without requiring any additional input function evaluations (Fig.~\ref{fig:remeshing}).  This process improves mesh quality while minimally impacting approximation accuracy. The process can be combined with rejection sampling, collapsing all rejected vertices. In combination with edge flipping this process generates meshes with even sizing and near perfect triangle shape. 

\paragraph{Curvature Adaptive Sampling.} 
As Fig.~\ref{fig:curvature} shows increasing refinement iteration enables us to support more adaptive sampling with more samples in higher curvature areas compared to flatter ones.

\paragraph{Implicits with Upper/Lower Distance Bounds.}
Our method can be sped up if the functions processed are not pure (discontinuous) OFs. If the function has an upper bound, as is the case with many neural functions which smoothly approximate the zero-level set in its immediate vicinity, we can eliminate the barrier test (Sec.~\ref{sec:refinement}) and simply disallow scaffold vertices within $\epsilon$-distance from the zero-level set. This speeds up computation and improves sampling quality. Given a lower bound, our binary search can be replaced by sphere tracing, and the barrier check can be similarly eliminated, significantly speeding computation.  

\section{Conclusion}
\label{sec:conclusion}  

We propose ADS, a novel method for sampling and iso-surfacing occupancy functions. Unlike prior iso-surfacing approaches for OFs, it generates random samplings. ADS requires significantly fewer function evaluations than all alternatives and is notably faster than prior random sampling methods. It is also simple to implement. 

\begin{figure*}
	\centering \includegraphics[width=\linewidth]{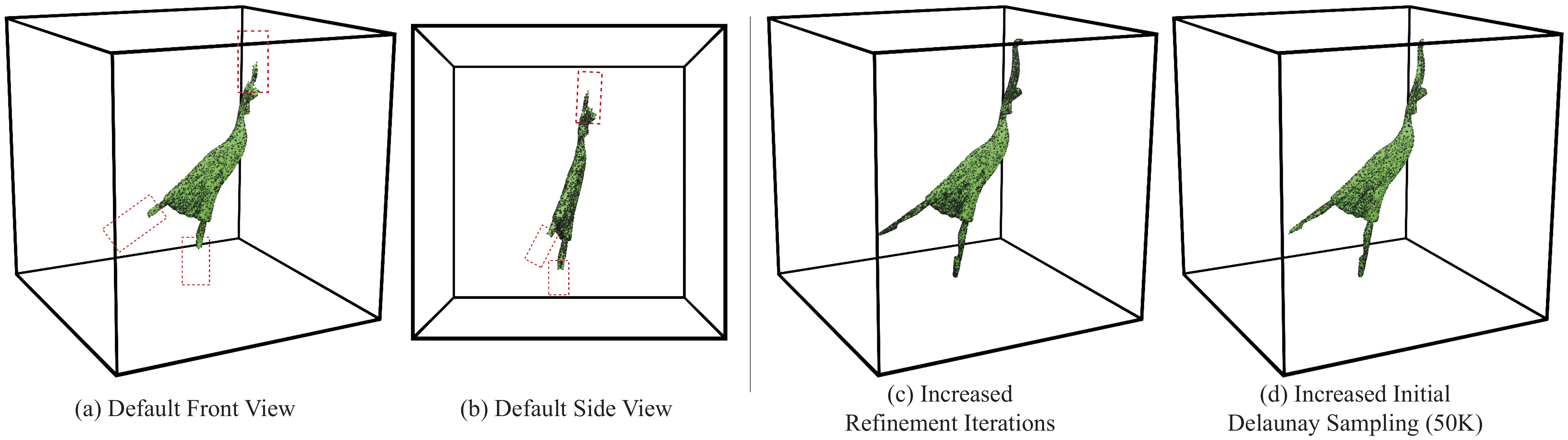}
	\caption{When the input OF depicts very thin shapes, using our default method and default initial scaffold may fail to capture them even at fine mesh resolution (a,b); this can be mitigated  by using a significantly higher number of refinement iterations (c) or a finer initial scaffold (d). Box delineates the OF domain. }
	\label{fig:limitations}
\end{figure*}

A core limitation of ADS, shared by other sampling strategies, is the need for the initial sampling to at least weakly correlate with OF surface dimensions. 
When the iso-surfaced shapes are drastically thinner than the function domain, our initial scaffold may place very few points in their interior, causing poor reconstruction (Fig.~\ref{fig:limitations}ab). This can be mitigated by  applying more refinement iterations (Fig.~\ref{fig:limitations}c) or using a finer initial scaffold (Fig.~\ref{fig:limitations}d, 50K instead of 10K vertices). Another limitation of ADS is the prevalence of clusters on the surface which persist, although our barrier test alleviates this issue (see Sec.~\ref{sec:edge_refinement}).

\begin{acks}
	We thank Daniel Cui for their help with proofreading. We acknowledge the support of the Natural Sciences and Engineering Research Council of Canada (NSERC) grant RGPIN-2024-03981. Finally, this work is supported in part by the Institute for Computing, Information and Cognitive Systems (ICICS) and Advanced Research Computing (ARC) at the University of British Columbia (UBC).
\end{acks}

\bibliographystyle{ACM-Reference-Format}
\bibliography{ref}

\newpage

\end{document}